\documentclass[a4paper]{spie}
\usepackage[]{graphicx, subcaption}
\usepackage{amsmath,amssymb}
\usepackage{hyperref}

\title{DCGANs for Realistic Breast Mass Augmentation in X-ray Mammography}
\author{Basel Alyafi \supit{a}, Oliver Diaz\supit{a,b}, Robert Marti\supit{a}
\skiplinehalf
\supit{a} Computer Vision and Robotics Institute, University of Girona, Spain\\
\supit{b} Center for Digital Medical Imaging. Parc Tauli Hospital Universitari. Institut d'Investigacio i Innovacio Parc Tauli (I3PT). Sabadell, Spain
}

\begin{document}
\maketitle
\begin{abstract}
Early detection of breast cancer has a major contribution to curability, and using mammographic images, this can be achieved non-invasively. Supervised deep learning, the dominant CADe tool currently, has played a great role in object detection in computer vision, but it suffers from a limiting property: the need of a large amount of labelled data. This becomes stricter when it comes to medical datasets which require high-cost and time-consuming annotations. Furthermore, medical datasets are usually imbalanced, a condition that often hinders classifiers performance. The aim of this paper is to learn the distribution of the minority class to synthesise new samples in order to improve lesion detection in mammography. Deep Convolutional Generative Adversarial Networks (DCGANs) can efficiently generate breast masses. They are trained on increasing-size subsets of one mammographic dataset and used to generate diverse and realistic breast masses. The effect of including the generated images and/or applying horizontal and vertical flipping is tested in an environment where a 1:10 imbalanced dataset of masses and normal tissue patches is classified by a fully-convolutional network. A maximum of $\sim0.09$ improvement of F1 score is reported by using DCGANs along with flipping augmentation over using the original images. We show that DCGANs can be used for synthesising photo-realistic breast mass patches with a considerable diversity. It is demonstrated that appending synthetic images in this environment, along with flipping, outperforms the traditional augmentation method of flipping solely, offering faster improvements as a function of the training set size.
\end{abstract}
\section{Introduction}
\label{sec:intro}
Breast cancer is the second deadliest cancer in women globally after lung cancer. This disease was the most frequently diagnosed cancer in 154 countries and the first cause of cancer death in women in 100 countries in 2018 \cite{cancer_stats}. Computer-aided detection (CADe) systems have been a good alternative for double reading strategies in breast cancer screening benefiting from recent advances in supervised deep learning to reduce false negative and false positive cases. Supervised deep learning tools, however, require large amounts of annotated data. Unfortunately, publicly-available medical datasets are usually small and imbalanced (cancer vs non-cancer) due to privacy issues and the high cost of expert annotations. Generative Adversarial Networks (GAN) \cite{GAN} have shown promising results in synthesising medical images \cite{liver_aug}. GAN consist of a network (called generator or G) that learns the distribution of the input data implicitly by the aid of another network (called discriminator or D) which, in turn, tries to learn to distinguish real among synthetic images. Deep Convolutional GAN (DCGAN) \cite{radford_DCGAN}, chosen due to training stability, are used to neutralise a wide range of non-pertinent sources of variance given that the dataset has enough examples. Thereafter, the synthetic breast mass images are used to augment an imbalanced dataset for improving the classification performance. All materials are available online for scientific use (\href{https://github.com/Basel1991/Projects/tree/master/master_thesis/code}{link}).

\section{Description of Purpose}
\label{sec:purpose}
To use DCGAN in order to synthesise realistic and diverse breast masses to augment unbalanced datasets in a classification problem. The ultimate goal is to improve the performance of CADe systems in breast mass detection tasks.
\section{Materials}
\label{sec:materials}
The dataset used in this work is OPTIMAM Mammography Image Database (OMI-DB) \cite{Optimam}. This database includes over 145,000 cases (over 2.4 million images) and comprises unprocessed and processed digital mammograms from the NHS Breast Screening Programme of the United Kingdom. A subset of this database was obtained comprising over 80,000 cases. In this dataset, there are images from four vendors, however, only images belonging to Hologic Selenia Dimensions (Hologic, Inc; Bedford, Massachusetts, USA) were used in this work. This database has expert annotations identifying the image and any clinical observation. A total of 2,215 mass lesion and 22,000 normal tissue patches were extracted with size $128 \times 128$ pixels after applying histogram normalisation. Negative patches (normal tissue) were extracted randomly given that there was no overlap with the background or masses.

\section{Method}
\label{se:methods}
DCGAN \cite{radford_DCGAN} was used in this work with an additional layer for both of G and D to allow the generation of $128 \times 128$ pixel patches. The aim of the generator is to learn the mapping between the latent space (the normal distribution in this case) and the space of breast mass in a sense that it can transform a 200-value vector from the latent space to a breast mass image that can fool the discriminator.
The loss functions used to train the DCGAN, originally recommended in Ref. \citenum{GAN}, were cross entropy for D (Equation (\ref{Ld})) and the non-saturated loss for G (Equation (\ref{Lg})).
\begin{align}
{L}_{D}&=  -E_{x\ \in\ P_x,\ z\ \in P_z}[\log(x) + \log(1-D(G(z)))], \label{Ld}\\
{L}_{G}&= -E_{z\ \in\ P_z}[\ log(D(G(z)))\ ] \label{Lg},
\end{align}
where $P_x$ is the distribution of real breast masses, and $P_z$ is the normal distribution with zero mean and unitary standard deviation. As mentioned in Ref. \citenum{GAN}, this $L_G$ is preferred to $log(1-D(G(z)))$ because it has larger gradients at the beginning of the training process which makes G learn faster. 
\begin{figure}[htb]
	\begin{subfigure}[t]{0.47\textwidth}
    \centering
    \includegraphics{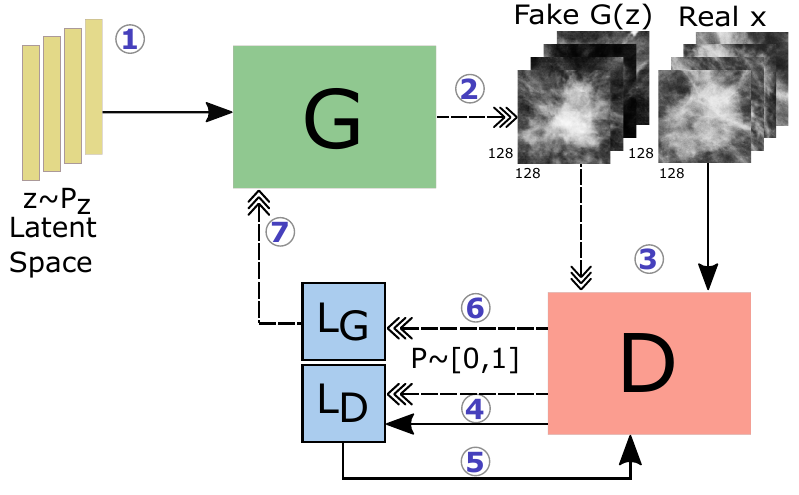}
    \caption{}
    \label{fig:GAN_training}
	\end{subfigure}
	\hspace{.2cm}
	\begin{subfigure}[t]{0.5\textwidth}
    \centering
    \includegraphics[keepaspectratio, scale= 0.65]{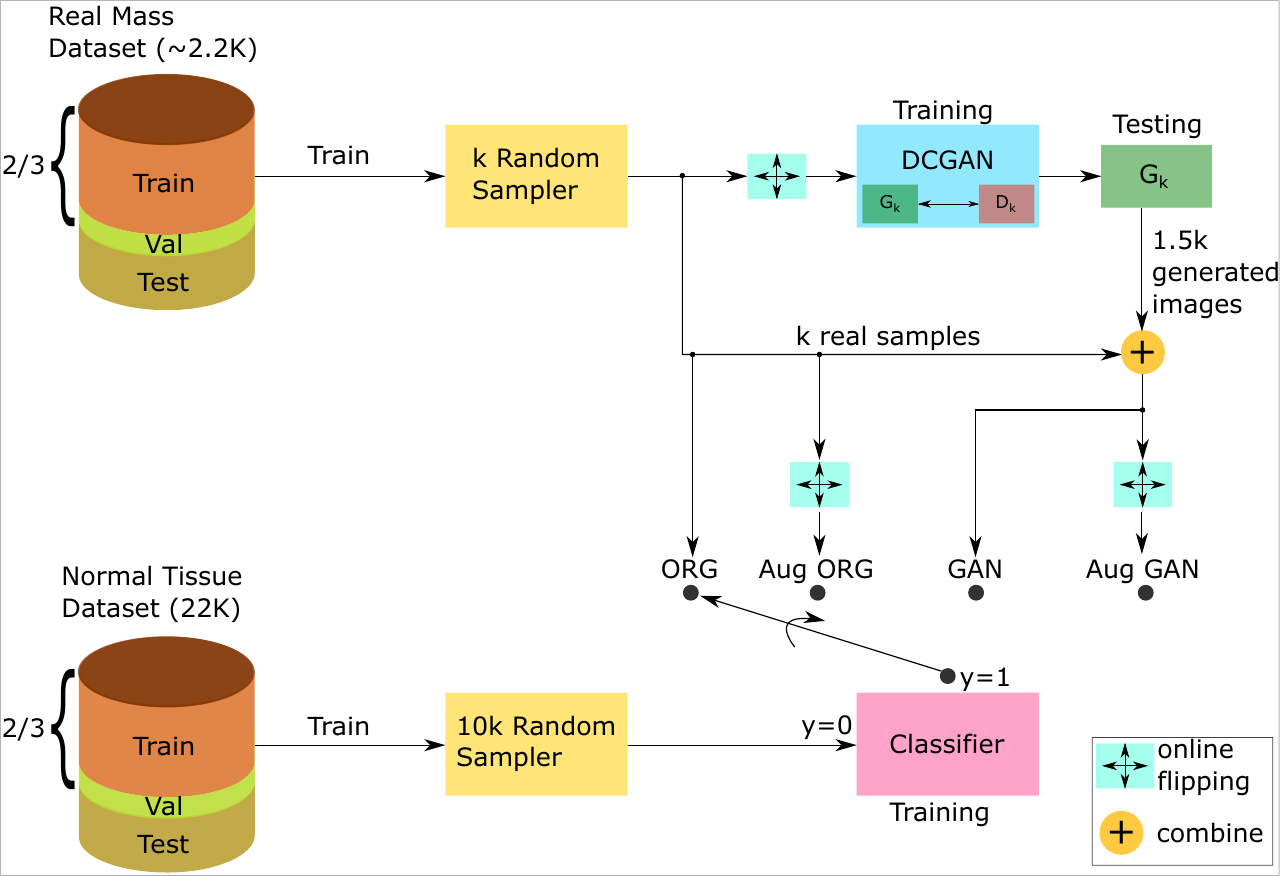}
    \caption{}
    \label{fig:methodology}
    \end{subfigure}
\caption{(a) Training DCGAN. Dotted arrows refer to fake patches. Steps from one to seven are: generate a noise batch z, forward z through G, forward the real and fake batches through D, calculate $L_D$, update D, calculate $L_G$, and update G. (b) The proposed framework for evaluating the DCGAN when used in data augmentation for supporting the minority class in an unbalanced dataset. Four strategies are investigated: ORG for using real images only, GAN for using real and synthetic images, Aug ORG for applying horizontal and vertical flipping on real images only, and Aug GAN for applying horizontal and vertical flipping on real and synthetic images.}
\end{figure}
Fig. \ref{fig:GAN_training} shows a schematic framework of the DCGAN. For each training iteration, 64 random latent vectors are sampled $z\in P_z;\ P_z=\mathcal{N}(0,1)$ (step 1 in Fig. \ref{fig:GAN_training}). This pure-noise batch is normalized to the range $[-1,1]$ then forwarded through G to generate a batch of fake images (G(z)) (step two).
These fake images are normalised to the range $[0,1]$ then forwarded through D to get realism probabilities, see step three with dashed arrows. An equal-size batch of real images is normalised and forwarded through D to learn the boundary between real and fake breast mass spaces, see step three with the dense arrow. In step four, Equation (\ref{Ld}) is used to calculate $L_D$, then D parameters are updated (step five). Thereafter, the fake batch is forwarded through D and Equation (\ref{Lg}) is used to calculate $L_G$ in step six. Backpropagation is done eventually to update G parameters (step seven). To complete one epoch, these steps are repeated until all the real breast mass patches are covered. 
As recommended in Ref. \citenum{GANtraining_techniques}, one-sided label smoothing was used to reduce over-confidence problems. In addition, conventional data augmentation, horizontal and vertical flipping, was used for increasing the diversity of the generated images. One critical issue was faced during training was the checkerboard effect in which a grid pattern appears in the synthesised images. The solution was inspired by a talk of Goodfellow \cite{nips2016}
, where the use of different kernel sizes between G and D was suggested.
In order to evaluate the trained generator, an augmentation environment was used where a 1:10 imbalanced dataset of masses (positive minority class) and normal tissue (negative majority class) was classified by a fully-convolutional network. In this setting, the classifier has a similar architecture to the DCGAN discriminator. Fig. \ref{fig:methodology} shows the pipeline used to evaluate the effect of data augmentation using four different approaches.
In addition, these augmentation effects on classification were investigated for different sizes of the training dataset.
For training the classifier, $60\%, 6.6\%$ and $33.4\%$ were used for training, validation and testing, respectively.
With respect to the positive class (2,215 breast masses), the training part was divided as \{$P_k; k \in \{100, 250, 500, 750, 1000, 1300\}$\}, where the subscripts refer to the size of the training subset. These subsets were sampled so that each subset is contained in the next larger one. Regarding the negative class (22,000 normal tissue patches), training subsets were designed to have an imbalance ratio of 10 \{$N_{1000}, N_{2500}, N_{5000}, N_{7500}, N_{10000}, N_{13000}$\}. 
The four augmentation approaches investigated (see Fig. \ref{fig:methodology}) are:
\begin{itemize}
    \item \textit{ORG}: using original images, the input for the classifier is $P_k$ as positive images plus $N_k$ as negative.
    \item \textit{Aug ORG}: original images were augmented using random horizontal and vertical flipping.
    \item \textit{GAN}: the training set of the classifier is $k$ real masses and $1.5 \times k$ synthetic masses as the positive class, and $10 \times k$ normal tissue patches as the negative class.
    \item \textit{Aug GAN}: the $1.5\times k$ generated images as well as the real ones were augmented on the fly by random horizontal and vertical flipping. 
    \end{itemize}
Because the dataset is imbalanced, F1 score was used as an evaluation metric. This provides equal importance to precision and recall. As observed in Fig. \ref{fig:methodology}, the test and validation sets were fixed for all k's. 3-fold cross validation was used to assure reliable results.

\section{Results and Discussion}
\label{sec:results}
\begin{figure}[htb]
	\begin{subfigure}[t]{0.45\textwidth}
	\centering
	\includegraphics[scale=0.85]{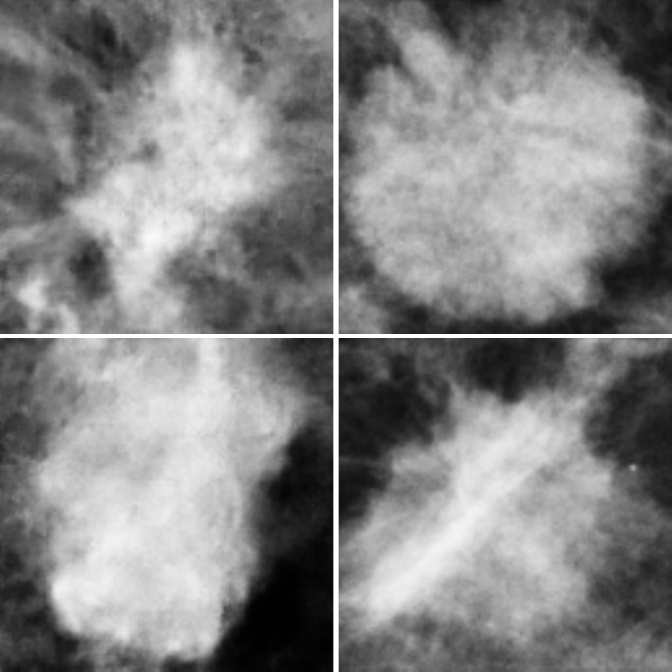}
	\caption{}
	\label{fig:synthetic_samples}
	\end{subfigure}
    \hspace{0.5cm}
    \begin{subfigure}[t]{.45\textwidth}
    \centering
    \includegraphics[scale=1]{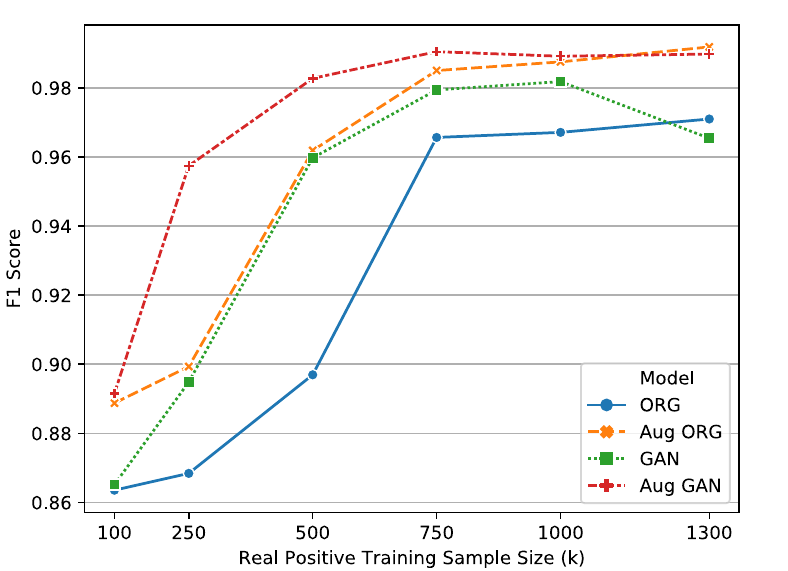}
    \caption{}
    \label{fig:F1Score_augmentation}
    \end{subfigure}
    \caption{(a) Synthetic (left column) and real (right column) breast mass lesions. (b) F1 score as a function of the size of the positive training set investigating four approaches: ORG, GAN, Aug ORG, and Aug GAN.}
\end{figure}

Fig. \ref{fig:synthetic_samples} shows two synthetic masses (left column) and two real masses (right column) depicting that DCGAN could generate visually-similar masses to the ones it was trained on. Moreover, Fig. \ref{fig:F1Score_augmentation} shows the F1 score for different training sizes, where each line represents one augmentation approach. The blue line (\textit{ORG}) shows the classification results using the original unbalanced training dataset. As more training images are available, the classifier increases its performance until k=750 where the performance saturates. When comparing the blue and the green lines (\textit{GAN}), the latter shows faster improvements which shows that the generator has learned to unlock unseen images in the real distribution helping the classifier to distinguish masses among normal tissue. If, on the other hand, the original data is augmented using horizontal and vertical flipping (the orange line of \textit{Aug ORG}), the classifier performs similarly to GAN (green) at medium sizes. Finally, the red line (\textit{Aug GAN}) shows the F1 score when random online flipping was applied on the combined real and synthetically-generated images. As can be depicted in the figure, \textit{Aug GAN} outperformed all other modes at any k, except a negligible drop at 1,300, with the maximum improvement at 250 with about 0.09 over ORG approach.

\section{Breakthrough work}
\label{sec:breakthrough_work}
The main contribution of this work is the development of a DCGAN-based model to generate synthetic breast masses.  Additionally, the performance of a fully-convolutional network classifier in an imbalanced mammography image dataset was investigated when the training dataset was enriched by including synthetic images, by flipping augmentation, or by a combination of them. This analysis on this scale is the first to the best of our knowledge. This work neither is being nor has been submitted to elsewhere.
\section{Conclusions}
\label{sec:conclusion}
In this study, we used a modified version of DCGAN to generate realistic breast mass patches with dimensions of $128 \times 128$ pixels. These synthetically-generated images were used to increase the size of the training dataset in a breast mass classifier. This was compared with conventional augmentation, i.e. flipping. Results, based on F1 score, suggest that classifiers with a training dataset smaller than 750 cases can greatly benefit from the synthetic images. On the contrary, conventional data augmentation strategies are sufficient for larger datasets.
\bibliography{spie}
\bibliographystyle{spiebib}
\end{document}